\documentclass[twocolumn, letterpaper, prb, footinbib, showkeys, showpacs, pdftex]{revtex4}
\usepackage[nodvipsnames]{color}
\usepackage{subfigure,amsmath}
\usepackage[pdftex]{graphicx}

\usepackage[%
	pdftex,%
	pdfstartview = {FitH},%
    pdftitle = {{Phase field simulations of coupled phase transformations in ferroelastic-ferroelastic nanocomposites}},%
	pdfsubject = {{We use phase-field simulations to study composites made of two different ferroelastic constituents (e.g. two types of martensite). The deformation of one material due to a phase transformation can elastically affect the other constituent and induce it to transform as well. We show that the phase transformation can then occur above its normal critical temperature, and even higher above this temperature in nanocomposites than in bulk composites. Microstructures depend on temperature, on the thickness of the layers, and on the crystal structure of the two constituents. Certain nanocomposites exhibit a great diversity of microstructures not found in bulk composites. Also, the periodicity of the martensite twins may vary over an order of magnitude based on geometry.}},%
	pdfauthor = {{Mathieu Bouville and Rajeev Ahluwalia}},%
	pdfkeywords = {{Ginzburg-Landau, multiferroics, nanostructure, phase-field model, shape-memory alloy}},%
]{hyperref}

\begin{document}

\title{Phase field simulations of coupled phase transformations\\in ferroelastic--ferroelastic nanocomposites}

\author{Mathieu Bouville}
	\altaffiliation{Current address: Department of Materials Science and Metallurgy, University of Cambridge}
	\email{mathieu.bouville@gmail.com}
	\affiliation{Institute of Materials Research and Engineering,\\ \mbox{A*STAR (Agency for Science, Technology and Research), Singapore 117602}}
	\affiliation{Materials Theory and Simulation Laboratory, Institute of High Performance Computing, A*STAR (Agency for Science, Technology and Research), Singapore 138632}

\author{Rajeev Ahluwalia}
	\email{rajeev@ihpc.a-star.edu.sg}
	\affiliation{Institute of Materials Research and Engineering,\\ \mbox{A*STAR (Agency for Science, Technology and Research), Singapore 117602}}
	\affiliation{Materials Theory and Simulation Laboratory, Institute of High Performance Computing, A*STAR (Agency for Science, Technology and Research), Singapore 138632}

\begin{abstract}
We use phase-field simulations to study composites made of two different ferroelastics (e.g.\ two types of martensite). The deformation of one material due to a phase transformation can elas\-ti\-cal\-ly affect the other constituent and induce it to transform as well. We show that the phase transformation can then occur above its normal critical temperature, and even higher above this temperature in nanocomposites than in bulk composites. Microstructures depend on temperature, on the thickness of the layers, and on the crystal structure of the two constituents~--- certain nanocomposites exhibit a great diversity of microstructures not found in bulk composites. Also, the periodicity of the martensite twins may vary over an order of magnitude based on geometry.
\end{abstract}
\maketitle

\section{Introduction}
Ferroelastics are materials which exhibit spontaneous strain, due to a displacive (diffusionless) structural phase transformation below a critical temperature. Typically, it is possible to switch between the different crystallographic variants (e.g.\ tetragonal unit cell elongated along $x$, $y$, or $z$) by applying an external stress. In martensite-based shape-memory alloys for example this gives rise to unusual mechanical properties \mbox{---such} as shape memory (a deformed material recovers its original shape upon heating) and pseudoelasticity (a large deformation completely recovered when the load is removed)\mbox{---,} leading to many technological applications.
Ferroelasticity is also found in ferroelectric and ferro\-magnetic materials. For example, 90$^\circ$ domains in ferroelectrics significantly contribute to the electromechanical response in piezoelectric applications. Similarly, the motion of ferroelastic domain walls notably influences the magnetic field-induced actuation in magnetoelastic materials.\cite{Cui04}

Ferroelectric and magneto\-strictive materials can be combined in multiferroic composites: the deformation of a ferroelectric due to an electric field triggers a trans\-formation in the magneto\-strictive material, so that the composite can `convert' an electric field into mag\-ne\-ti\-za\-tion.\cite{Nan-02, Fiebig-05, Zhang-07, Zhao-05} Alternatively, ferroic materials may be graded, i.e.\ properties such as composition and strain are made inhomogeneous, which leads to novel properties.\cite{Ban-03} In both cases, the presence of different ferroelastic domains crucially influences the properties of the system.
The effective behavior of a composite made of two ferroelastics (different in terms of transition temperature, transformation strain, or crystal symmetry) will be influenced by the long-range elastic interactions due to misfit strains existing both within and between the two constituents.

In the present article we focus on long-range elastic interactions and how the ferroelasticity in one constituent of the composite may influence the microstructure and properties of the other,\cite{Zhao-05} particularly how it can trigger a phase transition in a material which would not otherwise transform (e.g.\ it is above its transition temperature). In particular we address the technologically important question of the enhancement of this coupling by tuning materials properties and the characteristic length scale of the composite, especially in the nanometer regime.
We use martensite--martensite composites as model systems. While composites made of martensite and polymers\cite{Bidaux-93} or ceramics\cite{Petrovic-92} have been studied, no work exists on martensite--martensite composites. Moreover, martensite has been studied mostly in the bulk form, with limited work at the nanoscale.\cite{Bouville_Ahluwalia-acta_mater-08}
\nocite{Bouville-PRL-06, Bouville-PRB-07}

\section{Square-to-rectangle martensite}
In this section we will study a system made of two ferroelastic materials that can undergo a square-to-rectangle transformation. In the next section, we will turn to the case of composites made of a ferroelastic material undergoing a square-to-rectangle transformation and a material transforming from square to rhombus. We will talk of `rectangle--rectangle composite' and `rectangle--rhombus composite' respectively.

\subsection{Phase-field model}
The free energy of our two-dimensional system is based on the usual non-linear elastic free energy density for a square-\linebreak[3]to-\linebreak[3]rectangle martensitic transition:~\cite{Falk80, Onuki-99, Bouville_Ahluwalia-acta_mater-08, Bouville-PRL-06, Bouville-PRB-07}
\begin{eqnarray}
	&G=\displaystyle\int \left\{ \dfrac{A_{22}}{2}\dfrac{T\!-T_m}{T_m}(e_2)^2 - \dfrac{A_{24}}{4}(e_2)^4 + \dfrac{A_{26}}{6}(e_2)^6 + \right. \nonumber\\
	&  \left. \dfrac{A_1}{2} \left[ e_1 - x\,(e_2)^2  \right]^2 + \dfrac{A_3}{2} (e_3)^2
+ \dfrac{k}{2} \|\nabla e_2\|^2 \right\} \mathrm{d}\mathbf{r}.
	\label{eq-G-e2}
\end{eqnarray}
\noindent $T$ is the temperature and $T_m$ the austenite--martensite transition temperature (it will be different in the two constituents of the composite). The coupling constant $x$ is related to the unit cell volume difference between austenite and martensite (even though our model is two-dimensional, we call $x$ a \emph{volume} change to follow common practice). We use values from Ref.~\onlinecite{Kartha95} for~$\{A_{ij}\}$.

In Eq.~(\ref{eq-G-e2}), $e_1$ is the hydrostatic strain, $e_2$ the deviatoric strain, and $e_3$ the shear strain:\cite{Falk80, Onuki-99, Bouville_Ahluwalia-acta_mater-08, Bouville-PRL-06, Bouville-PRB-07}
\begin{subequations}
\begin{align}
	e_1 &= (\varepsilon_{xx}+\varepsilon_{yy})/\sqrt{2},\\
	e_2 &= (\varepsilon_{xx}-\varepsilon_{yy})/\sqrt{2},\\
	e_3 &= \varepsilon_{xy}.
\end{align}
\end{subequations}
\noindent The $\{\varepsilon_{ij}\}$ are the linearized strain tensor components. For a square lattice (austenite) $e_2 = 0$, whereas $e_2 \ne 0$ for the two variants of rectangular martensite. The deviatoric strain $e_2$ is thus used to distinguish between austenite and rectangular martensite.
The strain gradient term $\nabla e_2$ in Eq.~(\ref{eq-G-e2}) does not contribute to the bulk energy, since there is no gradient within austenite or martensite; rather, it plays a role similar to interface energy (both austenite--martensite and martensite--martensite). This gradient prevents the system from creating an infinite number of interfaces.\cite{Khachaturyan-69}

To lowest order in $e_2$, the energy shown in Eq.~(\ref{eq-G-e2}) is quadratic with an extremum at $e_2=0$. At temperatures above $T_m$, the first term in Eq.~(\ref{eq-G-e2}) is positive, so that $e_2=0$ is a minimum, i.e.\ austenite is (meta)stable. On the other hand, if $T<T_m$ the first term in Eq.~(\ref{eq-G-e2}) is negative and $e_2=0$ is a maximum: austenite is unstable.

\begin{figure}
\centering
\includegraphics[width=8.5cm]{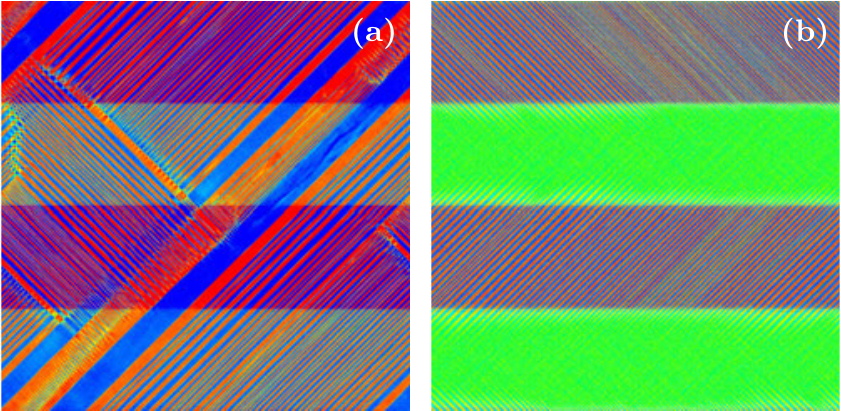}\vspace{-1mm}
\subfigure{ \label{mc-250K-0}}\vspace{-1mm}
\subfigure{ \label{mc-258K-0}}\vspace{-1mm}
	\caption{\label{mc-T-0}(Color online) Microstructures of bulk rectangle--rectangle composites at $x_a=0$ and (a)~250\,K and (b)~258\,K. Green: austenite, red: one (rectangular) martensite variant, and blue: the other (rectangular) martensite variant.}
\end{figure}

The evolution of the displacements is described by~\cite{landau-lifschitz}
\begin{equation}
	\rho \,\frac{\partial^2\, u_i(\mathbf{r}, t)}{\partial\, t^2} =
\sum_j\frac{\partial\, \sigma_{ij}(\mathbf{r}, t)}{\partial\, r_j} +
\eta \,\mathbf{\nabla}^2 v_i(\mathbf{r}, t),
\label{du_dt}
\end{equation}
\noindent where $\rho$ is a density, $\mathbf{v}$ is the time derivative of the displacements $\mathbf{u}$, and the stresses are given by
\begin{equation}
	\sigma_{ij}(\mathbf{r}, t) = \frac{\delta\, G}{\delta\, \varepsilon_{ij}(\mathbf{r}, t)},
\end{equation}
\noindent with $\delta$ the functional derivative. The second term on the right-hand side in Eq.~(\ref{du_dt}) is a viscous damping term; it is a simplification of the more general damping of Ref.~\onlinecite{Ahluwalia-acta_mater-06}.

\subsection{Description of the system}
\begin{figure}
\centering
\includegraphics[width=8.5cm]{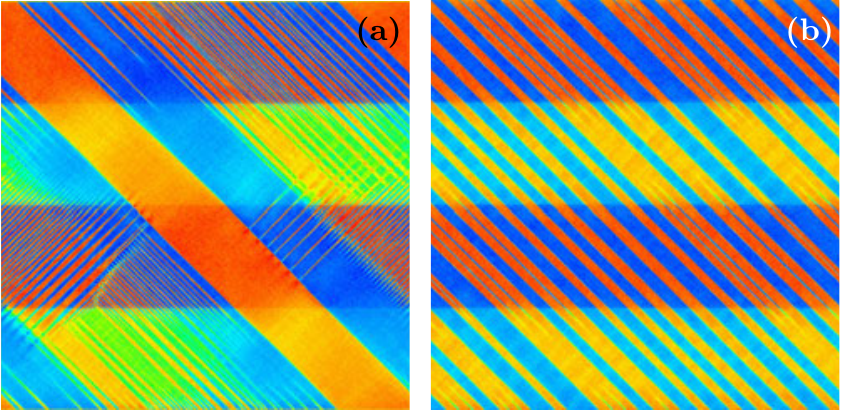}\vspace{-1mm}
\subfigure{ \label{mc-257K-0} }\vspace{-1mm}
\subfigure{ \label{mc-257K-05}}\vspace{-1mm}
\caption{\label{mc-257K-x}(Color online) Microstructures of bulk rectangle--rectangle composites at 257\,K and (a)~$x_a=0$ and (b)~$x_a=0.05$. Green: austenite, red: one (rectangular) martensite variant, and blue: the other (rectangular) martensite variant.}
\end{figure}

We consider a composite made of two materials that may form rectangular martensite; one has a transition temperature of $T_a=265$\,K, and $T_p=255$\,K for the other. At a temperature higher than both, the whole system remains austenitic; whereas below both, martensite is stable everywhere and the composite becomes completely martensitic, e.g.\ Fig.~\ref{mc-250K-0}. (In Fig.~\ref{mc-T-0} and all other pictures of microstructures, austenite is shown in green and the two martensite variants in red and blue\footnote{For interpretation of color in Figs.~\ref{mc-T-0}, \ref{mc-257K-x}, \ref{nc-200-258K-x}, \ref{nc-6x0-258xK-0}--\ref{sh-1000-T-0}, \ref{sh-thickness-258K-0}, \ref{transient}, and~\ref{sh-1200-253K-0-t} the reader is referred to the online version of this article.} \mbox{---red} corresponds to $e_2 > 0$, i.e.\ the variant with a unit cell elongated along~$x$, and blue to rectangles elongated along~$y$, $e_2 < 0$---, as in Ref.~\onlinecite{Bouville_Ahluwalia-acta_mater-08}.) We will therefore focus on intermediate temperatures, for which martensite is thermodynamically stable in only half the system. The other transformation may then occur only if it is triggered by the martensite already formed. For this reason we will call the two materials `active' and `passive' (the corresponding subscripts are $a$ and $p$, e.g.\ $x_a$ is the volume change in the active material).

The composite is made of four layers (from top to bottom: active--passive--active--passive), as seen in Fig.~\ref{mc-258K-0}. The system size is taken to be $6 \times 6~\mu$m (since the exact scale is material-dependent, we do not claim that our results are quantitatively correct for all systems) with periodic boundaries along both directions. The system, initially austenitic, is quenched to temperature $T$ and held at~it.

\subsection{Bulk composites}
Figures~\ref{mc-258K-0} and~\ref{mc-257K-0} show that a small difference of temperature can dramatically change the microstructure: at 257\,K the passive layers are almost fully martensitic as they would be below $T_p=255$\,K [e.g.\ Fig.~\ref{mc-250K-0}] whereas they remain austenitic at 258\,K. One can remark that at 257\,K martensite is thermodynamically unstable in the passive material~--- the phase transformation can take place there only thanks to the preceding martensite formation in the active layers.

\begin{figure}
\centering
	\includegraphics[width=8.6cm]{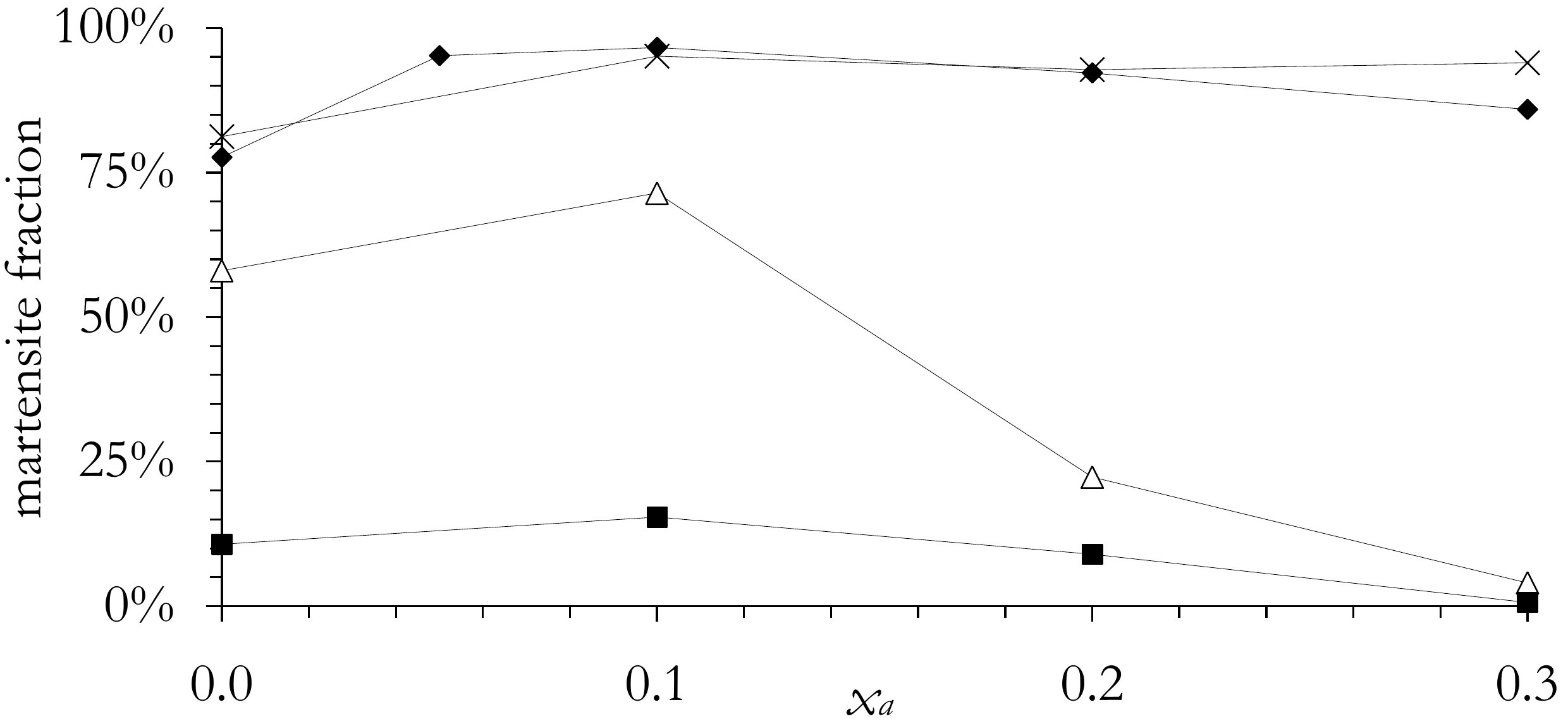}
	\caption{\label{Mfrac-bulk-rect}Martensite fraction in the passive layers of bulk rectangle--rectangle composites as a function of the volume change in the active layers $x_a$ at 256\,K (crosses), 257\,K (full diamonds), 257.5\,K (open triangles), and 258\,K (full squares).}
\end{figure}

If the martensite unit cell is bigger or smaller than the austenite unit cell then martensite formation generates hydrostatic strain. Such a situation will be taken into account by setting $x \ne 0$ in Eq.~(\ref{eq-G-e2}): if $x$ is different from zero then $e_2 \ne 0$ (i.e.\ martensite) leads to $e_1 \ne 0$.\cite{Bouville-PRL-06, Bouville-PRB-07, Bouville_Ahluwalia-acta_mater-08} Figures~\ref{mc-T-0} and~\ref{mc-257K-0} correspond to the case of $x = 0$. Figure~\ref{Mfrac-bulk-rect} shows the effect of a volume change in the active material. (We will limit ourselves to varying in different cases the value $x_a$ of the volume change in the active material~--- there will never be a volume change in the passive material, i.e.\ $x_p=0$.)
At 257\,K, the hydrostatic stress generated by the martensite of the active material favors martensite formation in the passive layers, Fig.~\ref{mc-257K-05}. Yet, Fig.~\ref{Mfrac-bulk-rect} shows that too high a value of $x_a$ hinders martensite formation in the active layer, which reduces the strain~--- there is a trade-off between strain generation and martensite formation in the active layer, which favors martensite formation in the passive layers at intermediate volume change. \cite{Bouville-PRL-06, Bouville-PRB-07}

\subsection{Nanocomposites}
Figures~\ref{mc-T-0}--\ref{Mfrac-bulk-rect} were obtained for bulk composites. In what follows we reduce the thickness of the layers to study nanocomposites. (The width remains set to 6~$\mu$m. To reduce the interaction between a layer and its own periodic images, the simulations for layers thinner than 250~nm have four pairs of layers rather than two.)

\begin{figure}
	\centering
	\includegraphics[width=8.6cm]{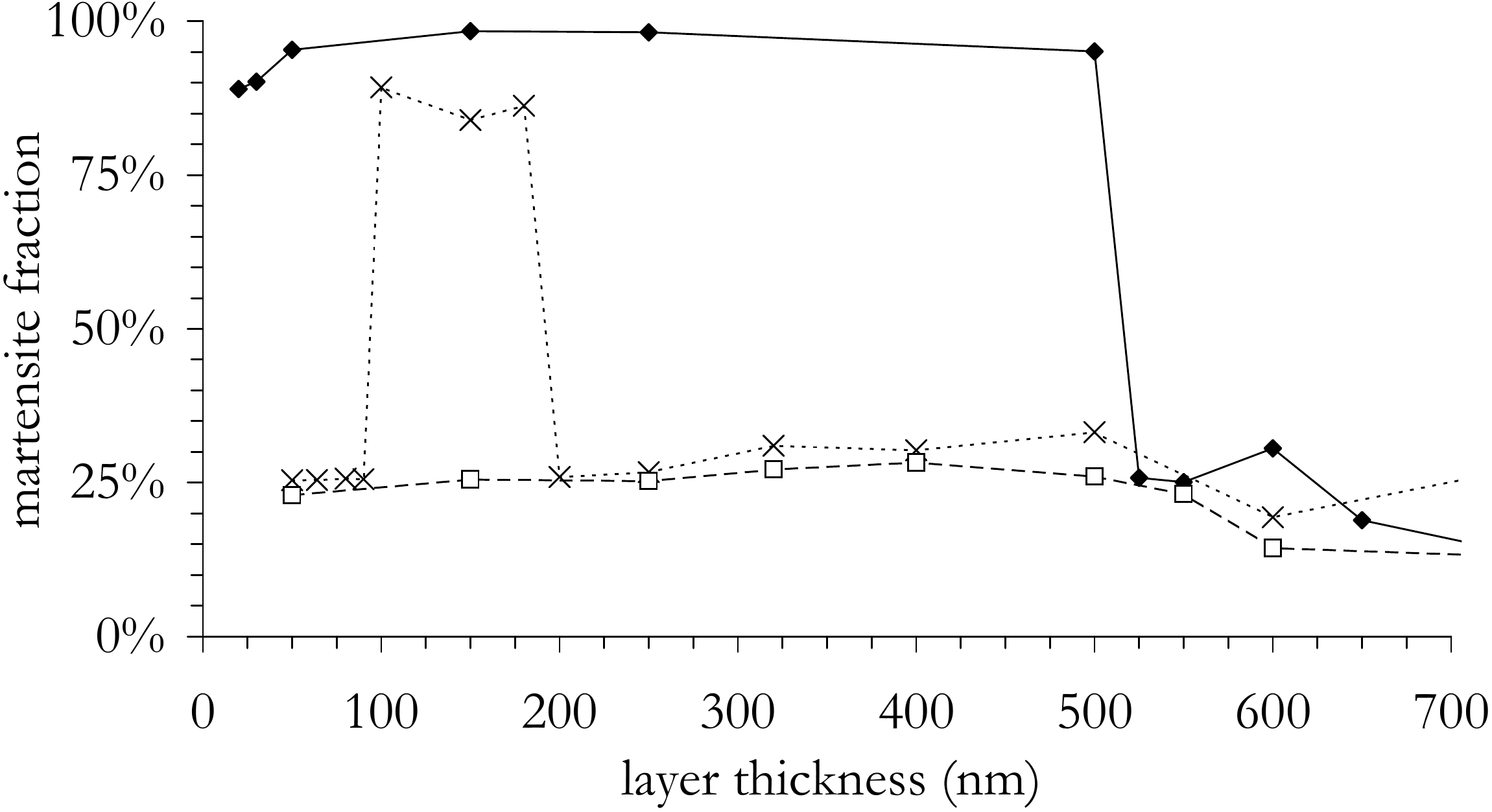}
	\caption{\label{Mfrac-thickness}Martensite fraction in the passive layers of rectangle--rectangle nanocomposites at 258\,K as a function of the layer thickness for different values of the volume change in the active layers: $x_a=0$ (closed symbols), $x_a=0.23$ (crosses), and $x_a=0.25$ (open symbols).}
\end{figure}

At 258\,K, martensite cannot form in the passive layers of bulk composites (with or without a volume change), as show Figs.~\ref{mc-258K-0} and~\ref{Mfrac-bulk-rect}. Figure~\ref{Mfrac-thickness} shows the martensite fraction in the passive layers as a function of their thickness and of the volume change in the active layers at 258\,K. It indicates that, even though hardly any martensite forms in the passive material of bulk composites at this temperature, martensite formation is possible in nanocomposites. (Also note that Fig.~\ref{Mfrac-thickness} confirms that 1.5~$\mu$m-thick layers are suitable to simulate `bulk' composites: they are at least three times as thick as layers exhibiting nanosize effects.)

\begin{figure}
\centering
\includegraphics[width=8.5cm]{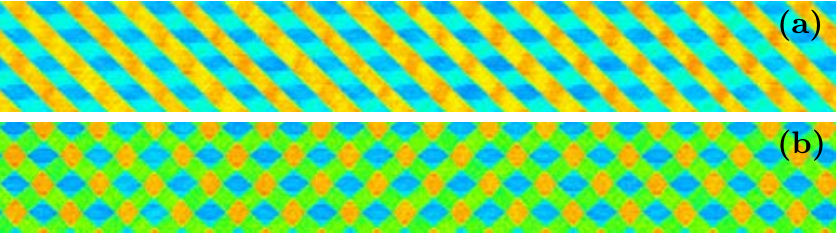}\vspace{-1mm}
\subfigure{ \label{nc-200-258K-23}}\vspace{-1mm}
\subfigure{	\label{nc-200-258K-24}}\vspace{-1mm}
	\caption{\label{nc-200-258K-x}(Color online) Microstructures of rectangle--rectangle nanocomposites with 100~nm-thick layers at 258\,K and (a)~$x_a=0.23$ and (b)~$x_a=0.24$. Green: austenite; red and blue: (rectangular) martensite.}
\end{figure}
One can notice a gap between about 25\% and 85\% martensite in Fig.~\ref{Mfrac-thickness}: when the passive layers are thin enough the microstructure switches from no or little martensite to pure martensite. For large volume changes, such as $x_a = 0.25$, there cannot be pure martensite at all at 258\,K, even for the thinnest layers. At intermediate $x_a$ (e.g.\ $x_a = 0.23$) there is an abrupt drop of the martensite fraction for very thin layers. A very small difference in $x_a$ or layer thickness can lead to a dramatic change of microstructure and of the volume fraction of martensite.
This is also quite obvious in Fig.~\ref{nc-200-258K-x}. One should point out that the `checkerboard' microstructure of Fig.~\ref{nc-200-258K-24} is not found in bulk composites. One can notice that its length scale is set by the thickness of the layers. Since there is no volume change in the passive layers the microstructure shown in Fig.~\ref{nc-200-258K-24} is due to the active layers: the passive layer `continues' the structure of the active layers, it does not impose its own microstructure.

\begin{figure}
\centering
\setlength{\unitlength}{1cm}
\begin{picture}(8.55, 8.7)(.2,0)
\shortstack[r]{
\\\vspace{-1.5mm}
\subfigure{
     \label{phase-diag-rect-x}
	\includegraphics[width=8.55cm]{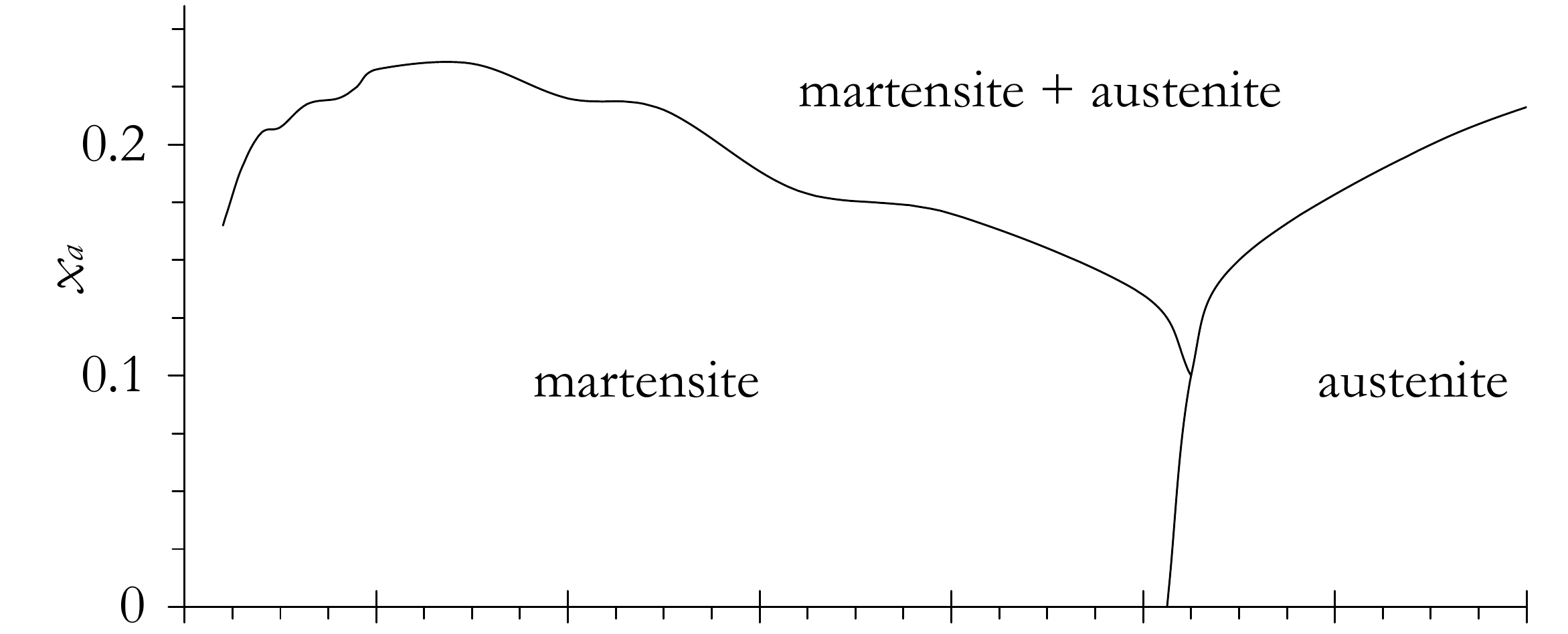}
	\put(-0.7, 3.1){(a)}
}\\\vspace{-1.5mm}
\subfigure{
     \label{phase-diag-rect-T}
	\includegraphics[width=8.55cm]{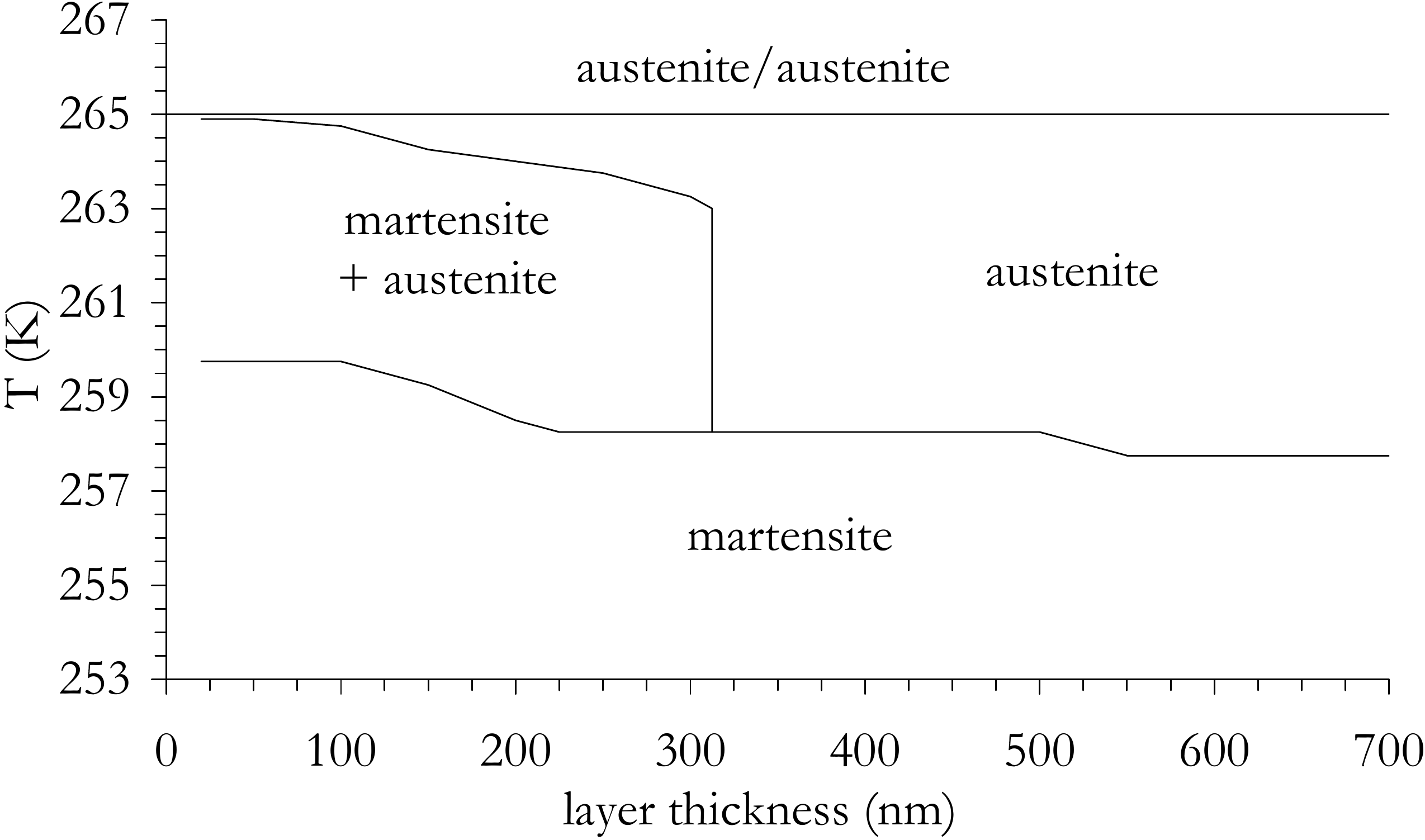}
	\put(-0.7, 4.6){(b)}
}}
\end{picture}
	\caption{Microstructure of the passive layers of rectangle--rectangle nanocomposites (a)~at 258\,K as a function of the layer thickness and of $x_a$, the volume change in the active layers and (b)~at $x_a=0$ as a function of the layer thickness and of the temperature. `Austenite/austenite' means that both active and passive layers are austenitic; for other cases, the active layer is martensitic and only the microstructure of the passive layer is shown. The `martensite' microstructure is shown in Fig.~\ref{nc-600-258K-0}, the `martensite + austenite' microstructure is exemplified by Fig.~\ref{nc-600-2585K-0}, and `austenite' by Fig.~\ref{nc-650-2585K-0}.}
\end{figure}

Figure~\ref{phase-diag-rect-x} gives the microstructure of the passive layers as a function of the layer thickness and of the volume change. The thicker layers behave like the bulk: the passive layers remain austenitic. The microstructure of thinner layers depends on the volume change $x_a$: for low values of $x_a$ the passive layers are martensitic, whereas for larger volume changes ($x_a > 0.23$) there is a mixture of austenite and martensite (note that $x_a = 0.23$ plays no special role in bulk composites). These trends are similar to what was observed in Ref.~\onlinecite{Bouville_Ahluwalia-acta_mater-08}. In the presence of a volume change martensite formation generates stress, so that for a high enough value of $x_a$ a purely martensitic system is higher in energy than a martensite--austenite mixture.\cite{Bouville_Ahluwalia-acta_mater-08, Bouville-PRL-06, Bouville-PRB-07}
One can see a range of values of $x_a$ around 0.2 for which there are four regimes for increasing layer thickness, including two with a mixture of martensite and austenite.

\begin{figure}
\centering
\includegraphics[width=8.5cm]{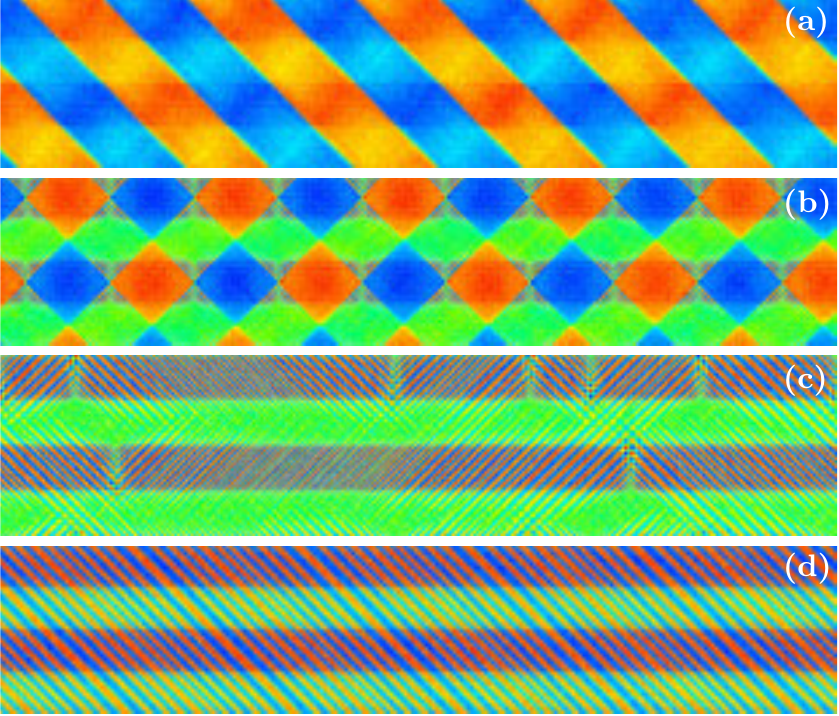}\vspace{-1mm}
\subfigure{	\label{nc-600-258K-0}}\vspace{-1mm}
\subfigure{	\label{nc-600-2585K-0}}\vspace{-1mm}
\subfigure{	\label{nc-650-2585K-0}}\vspace{-1mm}
\subfigure{	\label{nc-600-257K-0}}\vspace{-1mm}
	\caption{\label{nc-6x0-258xK-0}(Color online) Microstructures of rectangle--rectangle nanocomposites at $x_a=0$. (a)~300~nm-thick layers at 258\,K, (b)~300~nm-thick layers at 258.5\,K, (c)~325~nm-thick layers at 258.5\,K, and (d)~300~nm-thick layers at 257\,K. Green: austenite; red and blue: (rectangular) martensite.}
\end{figure}

Above 265\,K, austenite is the ground state everywhere, so that no martensite forms (even in the active layers). Below 265\,K, the active material is martensitic and Fig.~\ref{phase-diag-rect-T} shows the microstructure in the passive layers of the rectangle--rectangle nanocomposites as a function of the layer thickness and of the temperature (in the absence of a volume change). At low temperature the passive layers are martensitic, e.g.\ Fig.~\ref{nc-600-258K-0}, this depends weakly on the thickness of the layers. At higher temperature, the thinner passive layers show a mixture of martensite and austenite [Fig.~\ref{nc-600-2585K-0}] and the thicker passive layers remain austenitic, as in Fig.~\ref{nc-650-2585K-0}. Figures~\ref{phase-diag-rect-T} and~\ref{nc-6x0-258xK-0} show that a small change in nanolayer thickness or in temperature around 300~nm and 258\,K can change the microstructure dramatically.%
\footnote{Some systems between 350 and 500~nm and between 258.5 and 260\,K exhibit a microstructure like that of Fig.~\ref{nc-600-2585K-0}, even though other systems with similar thickness and temperature are of the kind of Fig.~\ref{nc-650-2585K-0}. One must assume that the two microstructures are  close in energy and the transformation may or may not occur based on the details of the initial conditions.}

\begin{figure}
\centering
\includegraphics[width=8.5cm]{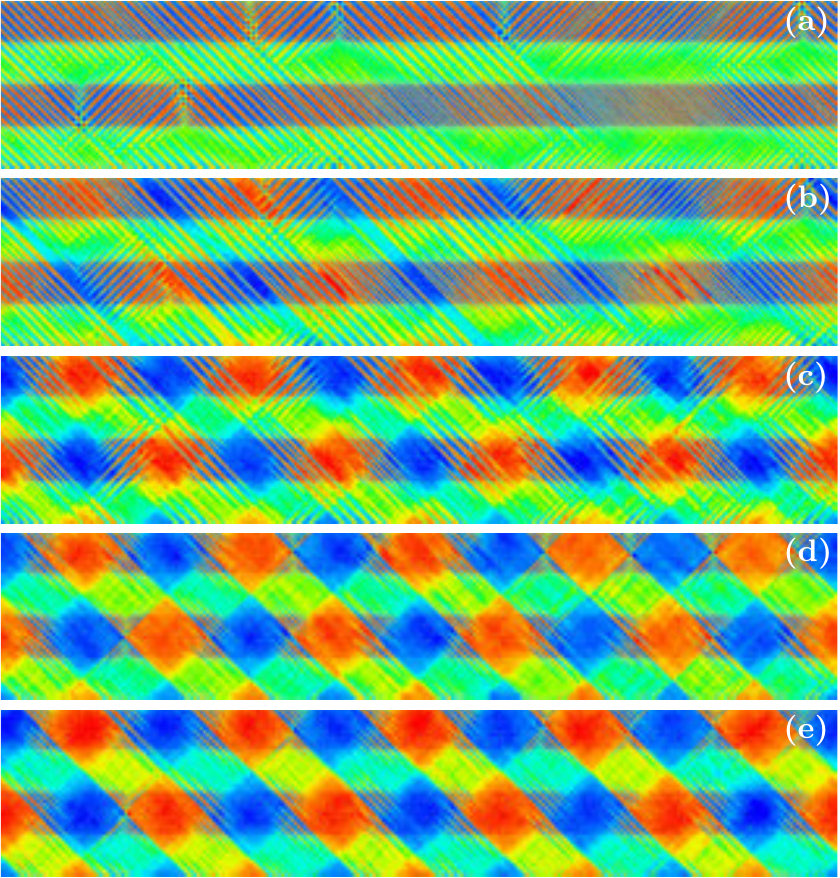}\vspace{-1mm}
\subfigure{	\label{nc-600-258K-0-t0050}}\vspace{-1mm}
\subfigure{	\label{nc-600-258K-0-t0060}}\vspace{-1mm}
\subfigure{	\label{nc-600-258K-0-t0065}}\vspace{-1mm}
\subfigure{	\label{nc-600-258K-0-t0070}}\vspace{-1mm}
\subfigure{	\label{nc-600-258K-0-t0075}}\vspace{-1mm}
	\caption{\label{nc-600-258K-0-t}(Color online) Time evolution of the microstructure of a rectangle--rectangle nanocomposite with 300~nm-thick layers at 258\,K and $x_a=0$. (a)~$t=50$ arbitrary units, (b)~$t=60$, (c)~$t=65$, (d)~$t=70$, and (e)~$t=75$. The final microstructure is shown in Fig.~\ref{nc-600-258K-0}. Green: austenite; red and blue: (rectangular) martensite.}
\end{figure}

The whole system is martensitic in both Figs.~\ref{nc-600-258K-0} and~\ref{nc-600-257K-0}; however, the period of the martensite twins is about an order of magnitude larger in Fig.~\ref{nc-600-258K-0}. Figure~\ref{nc-600-258K-0-t} shows the time evolution of the microstructure of a nanocomposite with 300~nm-thick layers at 258\,K, i.e.\ how the microstructure of Fig.~\ref{nc-600-258K-0} comes about. The early microstructure, Fig.~\ref{nc-600-258K-0-t0050}, is very similar to Fig.~\ref{nc-650-2585K-0}; it evolves to a state [Figs.~\ref{nc-600-258K-0-t0065} and~\ref{nc-600-258K-0-t0070}] very much like Fig.~\ref{nc-600-2585K-0}. In Fig.~\ref{nc-600-258K-0-t0060}, the ratio of the width of the red variant to the width of the blue variant varies spatially~--- this is already noticeable in Fig.~\ref{nc-600-258K-0-t0050}. Then certain twins disappear and the microstructure changes from twins of alternating martensite variants to large areas dominated by one variant, Fig.~\ref{nc-600-258K-0-t0065}. The active layers are then made of martensite `diamonds;' their width is equal to their height, so that the wavelength of the microstructure is set by the thickness of the layers rather than by the natural width of the twins [as is the case in Fig.~\ref{nc-600-258K-0-t0050}]. Between Figs.~\ref{nc-600-258K-0-t0070} and~\ref{nc-600-258K-0-t0075}, the austenite of the passive layers turns into martensite, leading to a purely martensitic system, Fig.~\ref{nc-600-258K-0}. (At higher temperature, e.g.\ Fig.~\ref{nc-600-2585K-0}, this latter transformation does not occur and the passive layers remain partly austenitic.)
This shows that these nanocomposites do not follow the well-known `square root law'\cite{Khachaturyan-69} where the periodicity of the twins is proportional to the square root of the plate width: the microstructure is set instead by the periodicity of the composite.
At lower temperature [Fig.~\ref{nc-600-257K-0}] the system is wholly martensitic; but unlike at 258\,K there was no intermediate austenite--martensite mixed microstructure, so the twins are much thinner.

\section{Square-to-rhombus martensite}

\subsection{Phase-field model}
We now turn to the case where the two materials making up the composite have different structures. The active material forms rectangular martensite below $T_a=265$\,K and the passive material forms rhombic martensite below $T_p=255$\,K. Again, we will focus on intermediate temperatures, for which rectangular martensite is thermodynamically stable but the rhombic one is not.

The material forming rectangular martensite (i.e.\ the active material) is governed by Eq.~(\ref{eq-G-e2}) and the square-to-rhombus martensite of the passive layers~by
\begin{eqnarray}
	&G=\displaystyle\int \left[ \dfrac{A_{32}}{2}\dfrac{T\!-T_p}{T_p}(e_3)^2 - \dfrac{A_{34}}{4}(e_3)^4 + \dfrac{A_{36}}{6}(e_3)^6 + \right. \nonumber\\
	&  \left. \dfrac{A_1}{2} (e_1)^2 + \dfrac{A_2}{2} (e_2)^2
+ \dfrac{k}{2} \|\nabla e_3\|^2 \right] \mathrm{d}\mathbf{r}.
	\label{eq-G-e3}
\end{eqnarray}
\noindent The rhombic martensite is then identified by the shear strain $e_3$ rather than by the deviatoric strain $e_2$.\footnote{In the figures we will plot the value of $e_2$ in the active layer and that of $e_3$ in the passive layer. Red thus corresponds to $e_2>0$ in the active material and to $e_3>0$ in the passive material; blue is $e_2<0$ in the active constituent and \mbox{$e_3<0$} in the passive one.} We set $A_2 = A_3$ and $A_{3i} = A_{2i}$. (Note that, since the rhombic martensite will always be in the passive layers and since we never have a volume change in the passive material, there is no coupling term $x$ in this equation.)

\begin{figure}
\centering
\includegraphics[width=8.5cm]{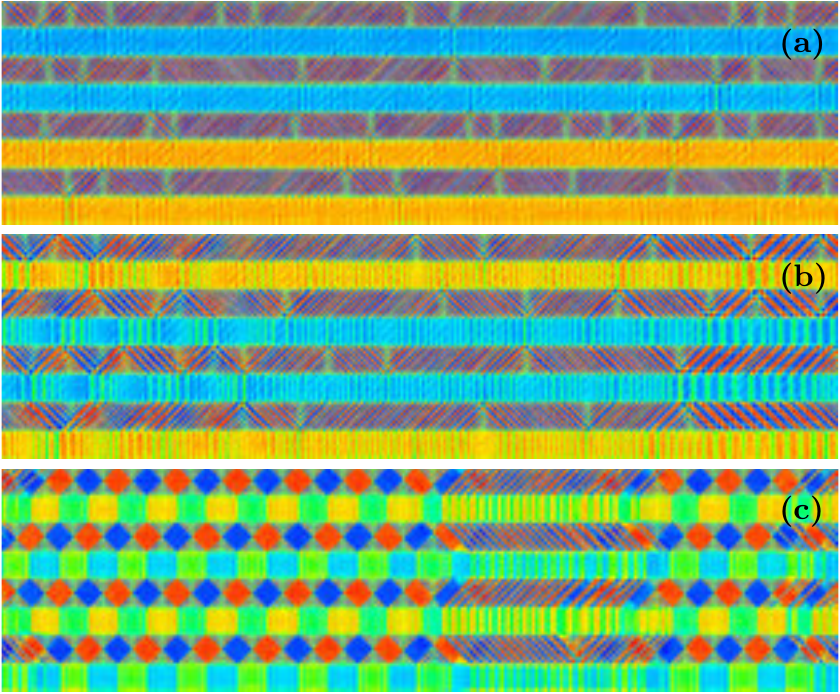}\vspace{-1mm}
\subfigure{ \label{sh-400-255K-0}}\vspace{-1mm}
\subfigure{ \label{sh-400-256K-0}}\vspace{-1mm}
\subfigure{ \label{sh-400-257K-0}}\vspace{-1mm}
	\caption{\label{sh-400-T-0}(Color online) Microstructures of rectangle--rhombus nanocomposites made of 200~nm-thick layers for $x_a=0$ at (a)~255\,K, (b)~256\,K, and (c)~257\,K. Green: austenite; red and blue: martensite.}
\end{figure}

\subsection{Nanocomposites}
Figure~\ref{sh-400-T-0} shows the microstructures of rectangle--rhombus nanocomposites made of 200~nm-thick layers for $x_a=0$ at various temperatures. At 255\,K, the passive layers are completely martensitic, with a single variant per layer, Fig.~\ref{sh-400-255K-0}. At 256\,K, there is also a single martensite variant per passive layer, but it alternates with small austenitic strips --- Fig.~\ref{sh-400-256K-0}. At 257\,K, Fig.~\ref{sh-400-257K-0}, every passive layer contains a single variant that alternates with austenite. At this temperature, one can notice an interaction between active layers: the `diamonds' in the active layers are aligned along [0\,1] with alternating $e_2>0$ [red in Fig.~\ref{sh-400-257K-0}] and $e_2<0$ (blue). This means that the passive layers influence the microstructure of the active layers. The `squares' of martensite in the passive layers are also aligned and they alternate between $e_3>0$ and~$e_3<0$.

\begin{figure}
\centering
\setlength{\unitlength}{1cm}
\begin{picture}(8.5, 5.9)(.2,0)
\shortstack[r]{
\\\vspace{-1.5mm}
\subfigure{
     \label{sh-1000-255K-0}
    \includegraphics[width=8.5cm]{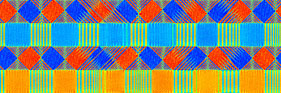}
	\put(-0.7, 2.4){\color{white}\bf(a)}
}\\\vspace{-1.5mm}
\subfigure{
     \label{sh-1000-2555K-0}
    \includegraphics[width=8.5cm]{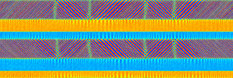}
	\put(-0.7, 2.4){\color{white}\bf(b)}
}}
\end{picture}
	\caption{\label{sh-1000-T-0}(Color online) Microstructures of rectangle--rhombus nanocomposites made of 500~nm-thick layers for $x_a=0$ at (a)~255\,K and (b)~255.5\,K. Green: austenite; red and blue: martensite.}
\end{figure}

Figure~\ref{sh-1000-T-0} shows the microstructures of rectangle--rhombus nanocomposites with 500~nm-thick layers (for \mbox{$x_a=0$).} At 255\,K, the passive layers alternate between one martensite variant and thin alternating twins of the two variants, Fig.~\ref{sh-400-255K-0}. This microstructure is different from that of Fig.~\ref{sh-400-257K-0} because it does not include any austenite. One can notice that the composite is periodic along one dimension in terms of its constituents but the pattern observed in Fig.~\ref{sh-1000-255K-0} [but also in Fig.~\ref{nc-600-2585K-0}] is periodic along both dimensions~--- such a spontaneous nanopatterning may have practical applications. At 256\,K, both martensite variants are found in a single passive layer, but the interface between them is parallel to the interface between layers, Fig.~\ref{sh-1000-2555K-0}.

\begin{figure}
\centering
\setlength{\unitlength}{1cm}
\begin{picture}(8.6, 11.5)(.2,0)
\shortstack[r]{
\\\vspace{-2mm}
\subfigure{
     \label{phase-diag-shear-T}
	\includegraphics[width=8.6cm]{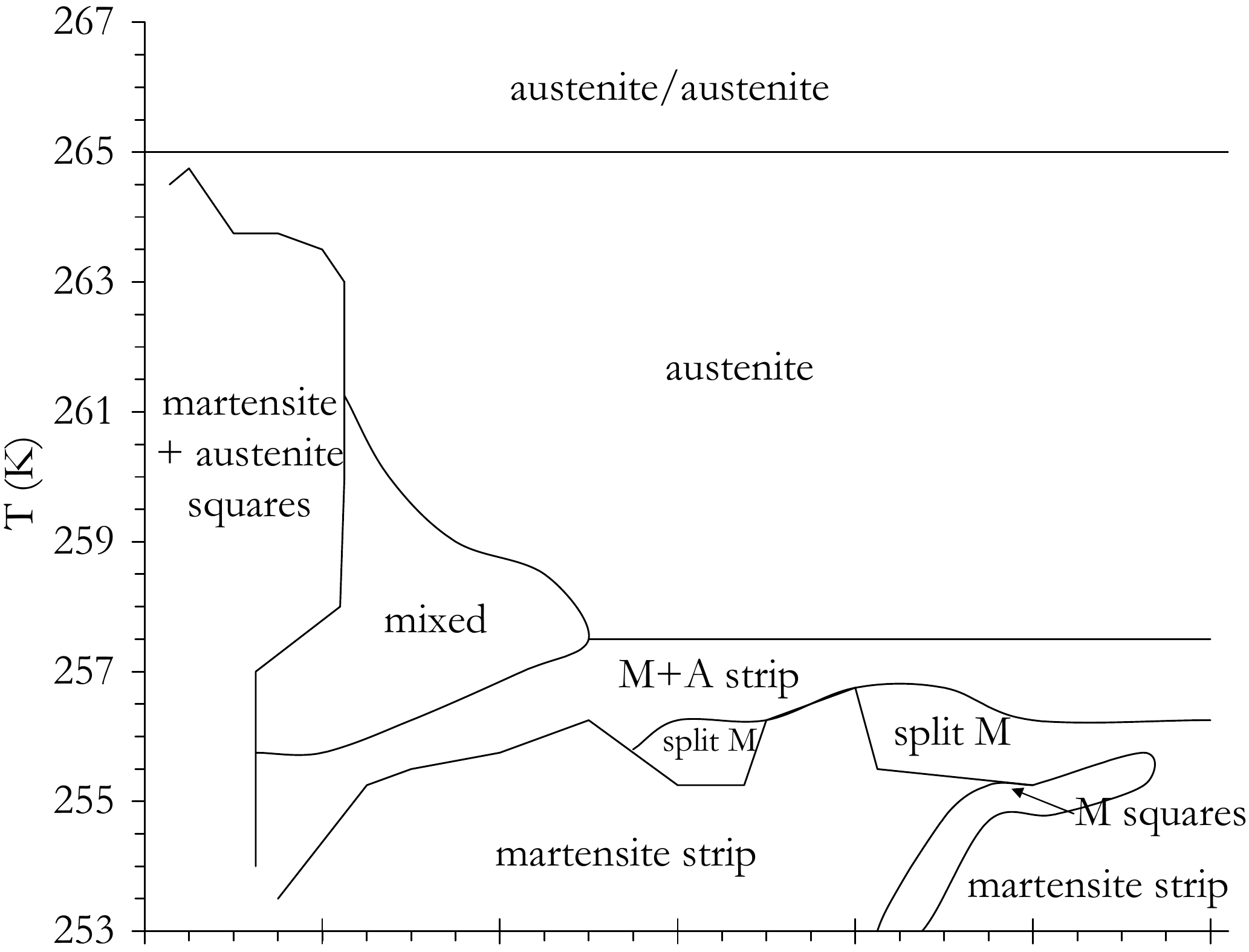}
	\put(-0.7, 6.){(a)}
}\\
\subfigure{
    \label{phase-diag-shear-x}
	\includegraphics[width=8.6cm]{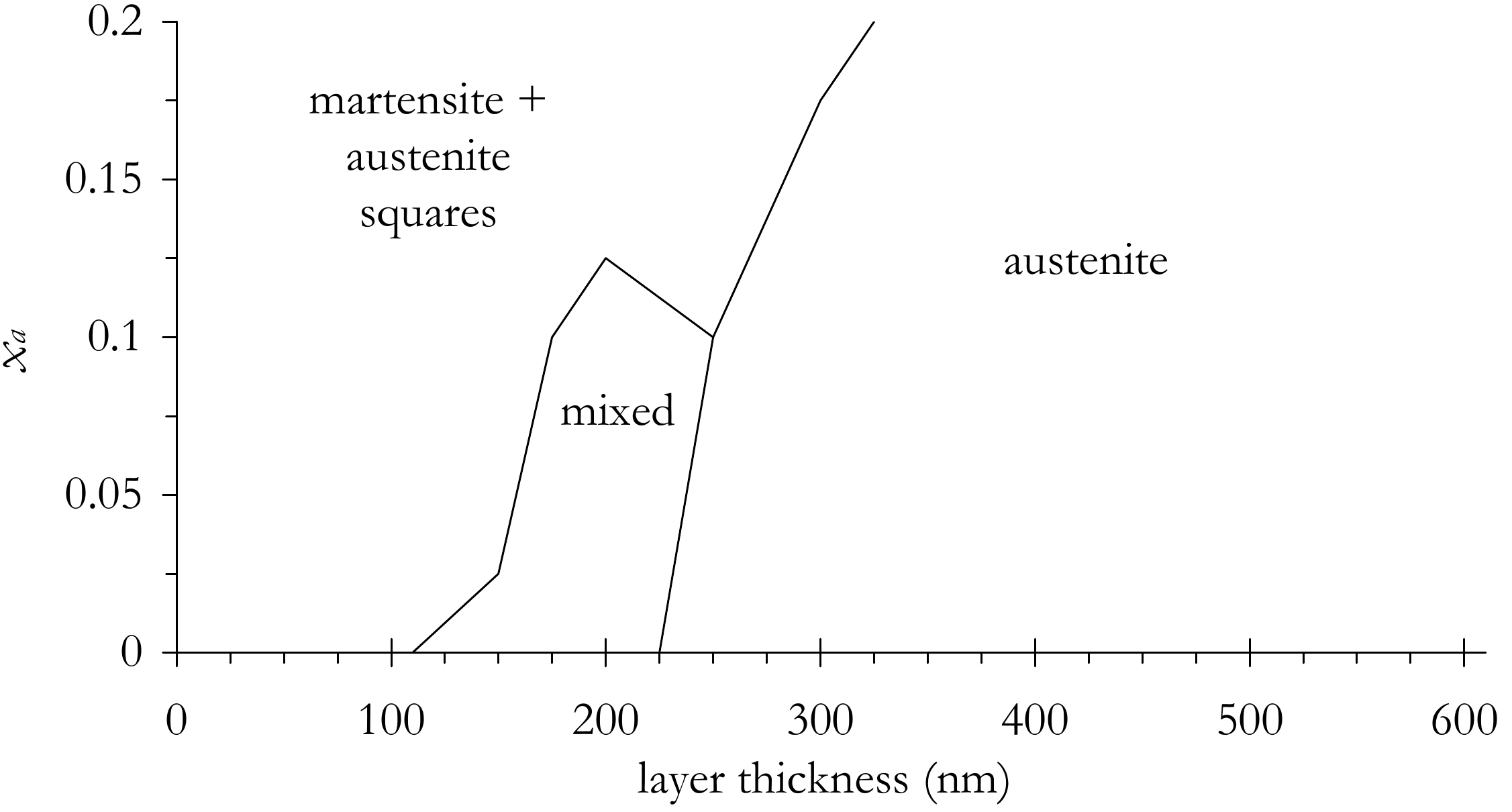}
	\put(-0.7, 4.){(b)}
}}
\end{picture}
	\caption{\label{phase-diag-shear}Microstructure of the passive layers of rectangle--rhombus nanocomposites (a)~at $x_a=0$ as a function of the layer thickness and of the temperature and (b)~at 258\,K as a function of the layer thickness and of $x_a$, the volume change in the active layers. `Austenite/austenite' means that both active and passive layers are austenitic; for other cases, the active layer is martensitic and only the microstructure of the passive layer is shown. The `martensite + austenite square' microstructure is exemplified by Fig.~\ref{sh-200-258K-0}, the `mixed' one by Fig.~\ref{sh-400-258K-0}, `austenite' by Fig.~\ref{sh-600-258K-0}, strips of martensite and austenite (`M+A strip')\ in Fig.~\ref{sh-400-256K-0}, strips of martensite (`martensite strip') in Fig.~\ref{sh-400-255K-0}, split strips of martensite (`split M') in Fig.~\ref{sh-1000-2555K-0}, and squares of one martensite variant alternating with squares made of twins of both variants (`M squares') are shown in Fig.~\ref{sh-1000-255K-0}.}
\end{figure}

\begin{figure}
\centering
\includegraphics[width=8.5cm]{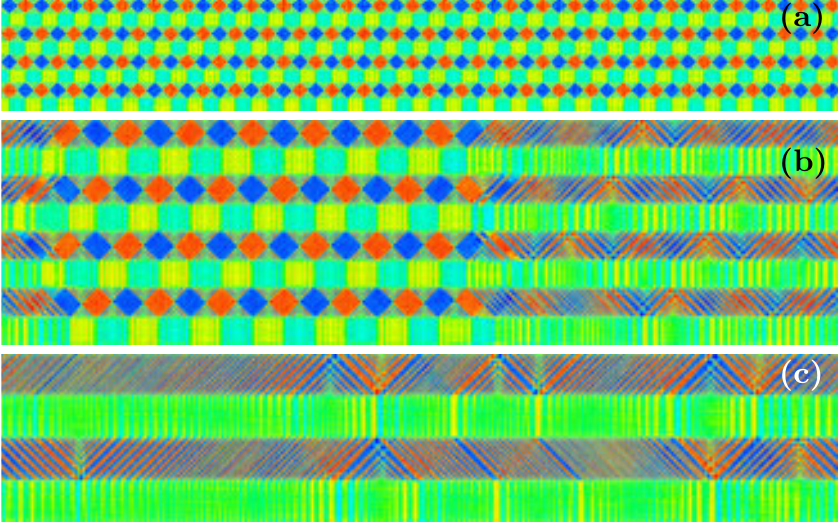}\vspace{-1mm}
\subfigure{ \label{sh-200-258K-0}}\vspace{-1mm}
\subfigure{ \label{sh-400-258K-0}}\vspace{-1mm}
\subfigure{ \label{sh-600-258K-0}}\vspace{-1mm}
	\caption{\label{sh-thickness-258K-0}(Color online) Microstructures of rectangle--rhombus nanocomposites at 258\,K for $x_a=0$ with layer thicknesses of (a)~100~nm, (b)~200~nm, and (c)~300~nm. Green: austenite; red and blue: martensite.}
\end{figure}

Figure~\ref{phase-diag-shear-T} gives the microstructure of the passive layers of rectangle--rhombus nanocomposites as a function of the layer thickness and of temperature. It is striking that the rectangle--rhombus nanocomposites exhibit a greater diversity of microstructures than the rectangle--rectangle ones, Fig.~\ref{phase-diag-rect-T}. Thin passive layers are made of squares of martensite and austenite, e.g.\ Fig.~\ref{sh-200-258K-0}, for all temperatures below 265\,K. Thicker layers are `mixed' as can be seen in Fig.~\ref{sh-400-258K-0}. Between about 257\,K and 265\,K, the thicker passive layers remain austenitic, as in Fig.~\ref{sh-600-258K-0} for instance. At lower temperature, one observes split layers of martensite [Fig.~\ref{sh-1000-2555K-0}]. At the lowest temperatures the passive layers turn to a single martensite variant, as in Fig.~\ref{sh-400-255K-0}, or to alternating `twins' of martensite and austenite [Fig.~\ref{sh-400-256K-0}]. These three microstructures are determined by temperature more than by layer thickness. At low temperature the larger systems form squares of one martensite variant alternating with squares made of twins of both variants, as shown in Fig.~\ref{sh-1000-255K-0}. 
One can remark in Fig.~\ref{phase-diag-shear-T} that the split martensite domain is itself split. Even though one expects the split layer to be higher in energy than a single variant, a symmetric configuration may be metastable because there no reason for the interface to move in one direction rather than the~other.

\subsection{Volume change}
\begin{figure}
\centering
	\includegraphics[width=8.6cm]{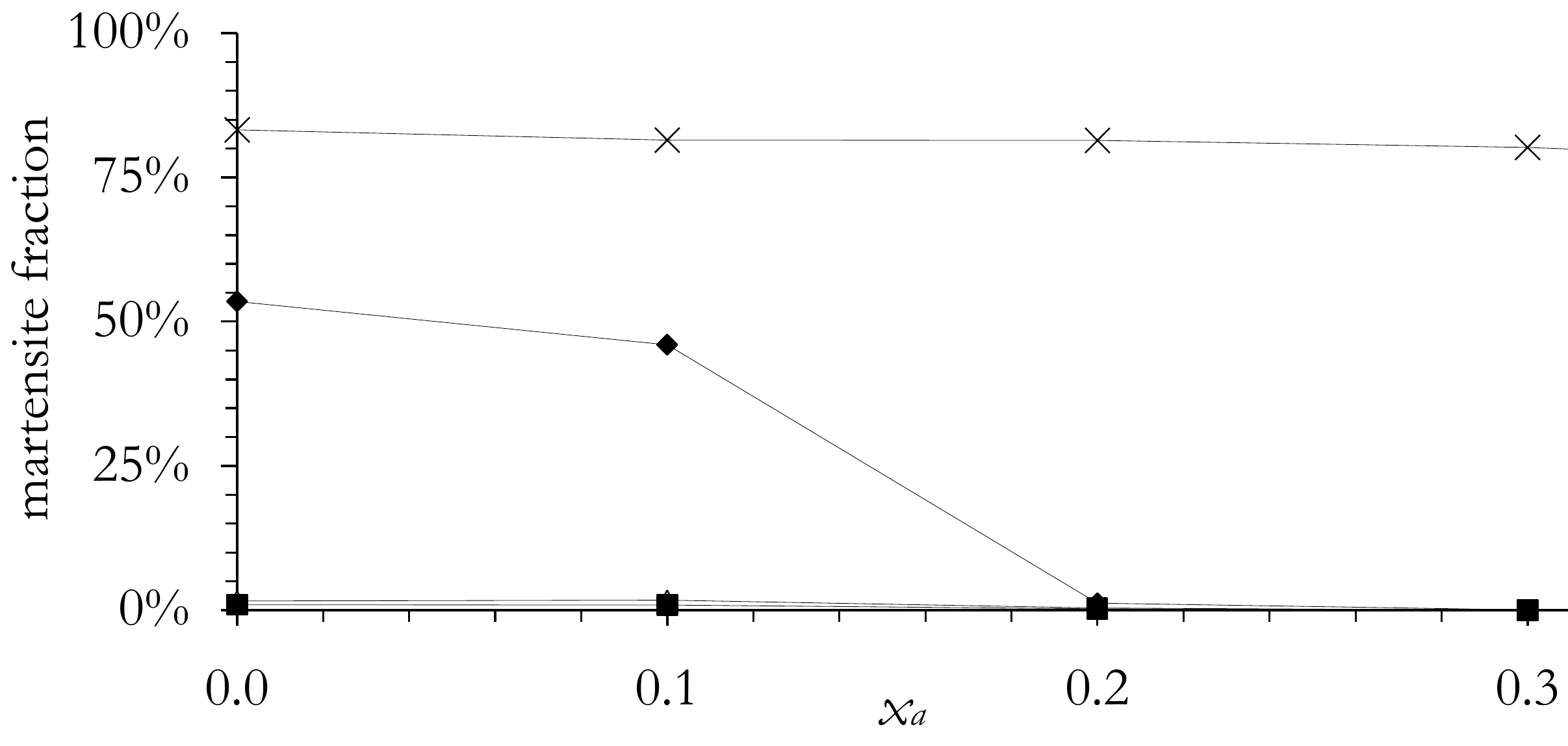}
	\caption{\label{Mfrac-bulk-shear}Martensite fraction in the passive layers of bulk rectangle--rhombus composites as a function of the volume change in the active layers $x_a$ at 256\,K (crosses), 257\,K (full diamonds), 257.5\,K (open triangles), and 258\,K (full squares).}
\end{figure}

Figure~\ref{Mfrac-bulk-shear} shows the effect of a volume change associated with the rectangular martensite (there is no volume change for the rhombic martensite) in bulk rectangle--rhombus composites. Unlike in the case of rectangle--rectangle composites (Fig.~\ref{Mfrac-bulk-rect}), at 257.5\,K the passive layers remain austenitic. In other words, the effective critical temperature for martensite formation is lower for the rectangle--rhombus system than for the rectangle--rectangle system.

Figure~\ref{phase-diag-shear-x} gives the microstructure of the passive layers as a function of the layer thickness and of the volume change. For thin layers the passive layers are made of squares of martensite, e.g.\ Fig.~\ref{sh-200-258K-0}. Thicker layers are `mixed' as can be seen in Fig.~\ref{sh-400-258K-0}. It seems that the complex microstructures found in Fig.~\ref{phase-diag-shear-T} at low temperature cannot exist at higher $T$ even in the presence of a volume change in the active layer.

\subsection{Interface orientation}
If we consider an interface (martensite--martensite or martensite--austenite) at an angle $\theta$ with the y-axis, we can determine what values of $\theta$ are allowed by elastic compatibility, which requires\cite{Kartha95}
\begin{equation}
	\frac{\partial^2 e_1}{\partial x^2} + \frac{\partial^2 e_1}{\partial y^2} - \frac{\partial^2 e_2}{\partial x^2} + \frac{\partial^2 e_2}{\partial y^2} - \frac{\partial^2 e_3}{\partial x\, \partial y}=0.
	\label{compatibility}
\end{equation}
\noindent For austenite and rectangular martensite the only contribution of $e_3$ to the free energy is through the $(e_3)^2$ term of Eq.~(\ref{eq-G-e2}); energy minimization with respect to $e_3$ then gives $e_3=0$. Likewise, in the absence of volume change, energy minimization gives $e_1=0$. Equation~(\ref{compatibility}) thus becomes
\begin{equation}
	-\frac{\partial^2 e_2}{\partial x^2} + \frac{\partial^2 e_2}{\partial y^2}=0.
	\label{compatibility-rect}
\end{equation}

Since the strain fields are one-dimensional (perpendicular to the interface), $e_1$, $e_2$, and $e_3$ depend only on $x \cos\theta + y \sin\theta$. Consequently, for $i=1,2,3$, one has\cite{Bouville-PRB-07}
\begin{equation}
	\frac{\partial^2 e_i}{\partial x^2} = e_i'' \cos^2\theta
	\text{\quad and\quad}
	\frac{\partial^2 e_i}{\partial y^2} = e_i'' \sin^2\theta.
\end{equation}
\noindent Equation~(\ref{compatibility-rect}) then gives $\cos 2\theta =0$. In order to satisfy compatibility, interfaces between two variants of rectangular martensite or between rectangular martensite and austenite must be aligned along $\langle 1\,1 \rangle$. Figures~\ref{nc-200-258K-x} and~\ref{nc-6x0-258xK-0} show good examples of both martensite--martensite and martensite--austenite interfaces along $\langle 1\,1 \rangle$, in both active and passive layers.

\begin{figure}
\centering
\setlength{\unitlength}{1cm}
\begin{picture}(8.5, 6.1)(.2,0)
\shortstack[r]{
\\\vspace{-2mm}
\subfigure{
     \label{sh-1000-2555K-0-t0060}
    \includegraphics[width=8.5cm]{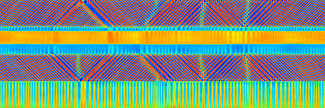}
	\put(-0.55, 2.45){\color{white}\bf(a)}
}\\\vspace{-2mm}
\subfigure{
    \label{sh-1100-2555K-0-t0110}
    \includegraphics[width=8.5cm]{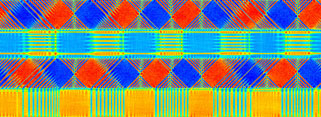}
	\put(-0.55, 2.65){\bf(b)}
}}
\end{picture}
	\caption{\label{transient}(Color online) Transient microstructures of rectangle--rhombus nanocomposites at 255.5\,K, $x_a=0$, and layer thicknesses of (a)~500~nm and (b)~550~nm. Green: austenite; red and blue: martensite.}
\end{figure}

\begin{figure}
\centering
\setlength{\unitlength}{1cm}
\begin{picture}(8.5, 7.1)(.2,0)
\shortstack[r]{
\\\vspace{-2mm}
\subfigure{
     \label{sh-1200-253K-0-t0050}
    \includegraphics[width=8.5cm]{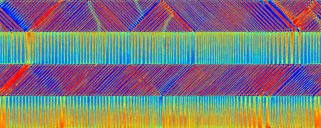}
	\put(-0.6, 3.){\color{white}\bf(a)}
}\\\vspace{-2mm}
\subfigure{
     \label{sh-1200-253K-0-t0060}
    \includegraphics[width=8.5cm]{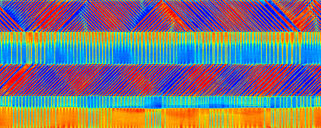}
	\put(-0.6, 3.){\color{white}\bf(b)}
}}
\end{picture}
	\caption{\label{sh-1200-253K-0-t}(Color online) Microstructure of a rectangle--rhombus nanocomposite with 600~nm-thick layers at 253\,K, $x_a=0$, and times (a)~$t=50$ arbitrary units and (b)~$t=60$. Green: austenite; red and blue: martensite.}
\end{figure}

In the case of rhombic martensite energy minimization gives $e_2=0$, so that only the last term in Eq.~(\ref{compatibility}) remains, leading to the condition: $\sin 2\theta =0$. In order to satisfy compatibility, interfaces between two variants of rhombic martensite or between rhombic martensite and austenite must therefore be along $\langle 1\,0 \rangle$. Figures~\ref{sh-400-T-0} and~\ref{sh-thickness-258K-0} for instance exhibit martensite--martensite and martensite--austenite interfaces along $[0\,1]$, while Fig.~\ref{sh-1000-2555K-0} shows an interface along $[1\,0]$.
Figure~\ref{sh-1000-2555K-0-t0060} shows different layers with qualitatively different microstructures~--- in particular, one passive layer has interfaces along $[0\,1]$ and the other along $[1\,0]$. (This microstructure is not stable; however, it is likely that actual systems, made of many layers, will exhibit different microstructures in different passive layers at equilibrium~--- especially close to the domain boundaries of Fig.~\ref{phase-diag-shear}.) Figure~\ref{sh-1100-2555K-0-t0110} shows a microstructure similar to Fig.~\ref{sh-1000-255K-0} except that there are martensite--martensite interfaces along both $[0\,1]$ and $[1\,0]$. Figure~\ref{sh-1200-253K-0-t} shows a microstructural evolution from martensite twins aligned along $[0\,1]$ to a microstructure we labeled `split martensite' with a martensite--martensite interface along $[1\,0]$; this `rotation' occurred without any martensite--martensite interface ever being along a direction other than $\langle 0\,1 \rangle$.

\section{Conclusion}
We used martensite--martensite composites as model systems to study the interaction between two ferroelastic materials, as in multiferroic composites for instance. We considered a composite made of two martensite-forming materials with different transition temperatures, so that at intermediate temperatures martensite is thermodynamically stable in only half the system. 

Our phase-field simulations showed that martensite can form in the passive layers of martensite--martensite composites even at temperatures at which it is not thermodynamically stable because, due to long-range elastic interactions, the transformation is triggered by the martensite already formed in the active layers. Martensite may form at even higher temperatures in nanocomposites than in bulk composites (for the narrowest layers it may even form as high in temperature as the transition temperature of the active material). However, this rise of the critical temperature may be weaker in the presence of a volume change in the active layers of the composites.

We simulated two kinds of systems: either both materials formed rectangular martensite or one martensite was rectangular and the other rhombic. These lead to different kinds of microstructures in the passive layers. In particular, the rectangle--rhombus nanocomposites exhibit a greater diversity of microstructures than the rectangle--rectangle ones. An interesting effect of using a given material in a composite is that the periodicity of the martensite twins may vary over an order of magnitude based on geometry. The properties of the microstructures  we observed in these ferroelastic--ferroelastic composites ---mechanical properties for martensite, polarization for ferroelectrics, etc.--- remain to be explored.

\bibliography{mart}
\end{document}